\documentclass[nocopyrights,sigplan,10pt,table,pdflatex]{acmart}

\usepackage{macros}
\usepackage{graphicx}
\usepackage[normalem]{ulem}

\usepackage{booktabs} 
\usepackage{amsmath}
\usepackage{wasysym}

\usepackage{natbib}

\newcommand{\vincent}[1]{} 
\newcommand{\cref}[1]{{\S\ref{#1}}}

\usepackage{collcell}
\usepackage{hhline}
\usepackage{pgf}
\usepackage{multirow}
\def\colorModel{hsb} 
\newcommand\ColCell[1]{
  \pgfmathparse{#1<100?1:0}  
    \ifnum\pgfmathresult=0\relax\color{white}\fi
  \pgfmathsetmacro\compA{1}      
  \pgfmathsetmacro\compB{#1/332}   
  \pgfmathsetmacro\compC{1}      
  \edef\x{\noexpand\centering\noexpand\cellcolor[\colorModel]{\compA,\compB,\compC}}\x #1
  } 
\newcolumntype{E}{>{\collectcell\ColCell}m{0.7cm}<{\endcollectcell}}  

\setcopyright{none}
\renewcommand\footnotetextcopyrightpermission[1]{} 
\newcommand\watermark[1]{%
    #1%
    \sbox0{#1}%
    \llap{%
    \makebox[\wd0][c]{
    \raisebox{.5\ht0}{
    \csname Gin@isotrue\endcsname
    \resizebox*{.8\ht0}{.8\ht0}{
    \parbox{10em}{
            \color{black!60}%
            {\sc techreport}
        }%
    }}}}%
}

\begin{document}
\date{}

\title{Evaluating the Red Belly Blockchain}

\author{Tyler Crain}
\affiliation{%
  \institution{University of Sydney}
}
\email{tyler.crain@sydney.edu.au}
\author{Christopher Natoli}
\affiliation{%
  \institution{University of Sydney}
}
\email{christopher.natoli@sydney.edu.au}
\author{Vincent Gramoli}
\affiliation{%
  \institution{University of Sydney}
  \institution{Data61-CSIRO}
}
\email{vincent.gramoli@sydney.edu.au}

\maketitle

\subsection*{Abstract}
{\small
In this paper, we present the most extensive evaluation of blockchain system to date.
To achieve scalability across servers in more than 10 countries located on 4 different continents, 
we drastically revisited Byzantine fault tolerant blockchains and verification of signatures.
The resulting blockchain, called the \emph{Red Belly Blockchain (RBBC)}, commits more than a hundred thousand transactions issued by permissionless nodes. These transactions are grouped into blocks within few seconds through a partially synchronous consensus run by permissioned nodes.
It prevents double spending by guaranteeing that a unique block is decided at any given index of the chain in a deterministic way by all participants. 

We compared the performance of RBBC against traditional Byzantine fault tolerant alternatives and more recent randomized solutions.
In the same geo-distributed environment with low-end machines, we noticed two interesting comparisons: \textit{(i)}~the RBBC throughput scales to hundreds of machines whereas the classic 3-step leader-based BFT state machine used by consortium blockchains cannot scale to 40 identically configured nodes;
\textit{(ii)}~RBBC guarantees transaction finality in 3 seconds and experiences a third of the latency that randomized-based solutions like HoneyBadgerBFT can offer. 
This empirical evaluation demonstrates that blockchain scalability can be achieved without sacrificing security.
}

\section{Introduction}

Blockchain systems~\cite{Nak08} aim at implementing a Byzantine fault tolerant replicated state machine (RSM) by totally ordering
blocks or sets of transactions that are issued by requesters. Various replicated state machines
have been proposed over the last decades to coordinate
servers, but they typically apply to a small set of replicas.
By contrast, blockchains aim at offering a peer-to-peer model where many geodistributed participants replicate the 
information and where many requesters can check their balance and issue cryptographically signed 
transactions. Permissioned (resp. permissionless) blockchains allow a pre-determined set of nodes (resp. all 
nodes) to be the deciders of new transaction blocks.
The limitations of existing blockchains, be they permissioned or permissionless, are their performance: 
the verification of all transactions is computationally intensive while reaching consensus is communication intensive.

In this paper, we evaluate a fast blockchain called the \emph{Red Belly Blockchain} 
(RBBC)\footnote{``Red belly'' is inspired by the name of a snake endemic to Sydney.} that is deterministic and does not assume synchrony.
RBBC offers a new sharding method that assigns, for each group of transactions, distinct groups of
\emph{proposer} and \emph{verifier} nodes.
\textit{(i)}~The sharding of proposers balances the communication load on multiple nodes, hence avoiding the congestion and slowdown induced by the least responsive node. 
As opposed to practical Byzantine consensus protocols that traditionally rely on a leader to propose a set of transactions, RBBC's multiple proposers combine distinct sets of transactions into a super block to commit more transactions per consensus instance.
\textit{(ii)}~The sharding of verifiers balances the computation load on different verifiers.
As opposed to existing blockchains whose nodes typically verify the same transactions, each of our transaction signature is verified by at least $t+1$ and at most $2t+1$ \emph{verifiers}
(where $t$ is the maximum number of faulty nodes). 

We conducted the most extensive evaluation of blockchain to date by evaluating \textit{(i)}~the peak throughput with the large bandwidth offered by hundreds of high end machines in a single data center, \textit{(ii)}~as compared to classic and randomized Byzantine tolerant blockchains, \textit{(iii)}~when attacked by a coalition of maliciously behaving machines and \textit{(iv)}~when deployed on low-end machines over 4 continents around the world.
While well-documented blockchain experiments already involved a large number of virtual machines~\cite{LNZ16,GHM17}, they typically spawn consensus participants in the same country, if not the same datacenter to make sure that message delays remain as low as possible.

Other blockchains even require that all message delays are lower than a bound imposed by the algorithm, an assumption called \emph{synchrony}~\cite{Nak08,Woo15,GHM17}, which can 
be difficult to achieve in practice and may be exploited to double spend~\cite{NG17,EGJ18}.
Some blockchain components that avoid the synchrony assumption cannot verify transaction signatures~\cite{SBV17}, 
allowing someone to withdraw from someone else's account. 
Randomized alternatives that terminate with high probability may require additional messages to implement a common coin~\cite{MSC16} that all participants can use.
Until now, purely Byzantine fault tolerant blockchains have been notoriously unscalable~\cite{Vuk15}, often reaching their peak throughput with 4 nodes~\cite{Buc16,SBV17}, hence tolerating at most one failure to maximize performance.
This even led companies to recently trade security for   
crash fault tolerance~\cite{ABB18}. 

For the sake of security, RBBC 
features the non Turing complete scripting language and the unspent transaction output (UTXO) model from Bitcoin~\cite{Nak08}. 
Each transaction is cryptographically signed and verified using Elliptic Curve Digital Signature Algorithm (ECDSA) keys. In contrast with Bitcoin, transactions are only verified by sufficiently many verifiers to cope with $t$ Byzantine nodes. The underlying consensus protocol run by $n$ nodes is especially designed to run in a blockchain over the Internet~\cite{CGLR18}. It is time optimal, resilience optimal in that it tolerates $t<\frac{n}{3}$ Byzantine nodes, and runs in a partially synchronous network.
To evaluate the performance of RBBC, we deployed it on up to 1000 nodes on up to 14 datacenters 
in 4 different continents. For most workloads, we tested the performance of RBBC 
across continents and observed tens of thousands of transactions committed per second.
The performance of our system peaks at 660,000 transactions per second, when run on 260 machines in a single datacenter with a low fault tolerance parameter $t$.

As cryptography is necessary but not sufficient to guarantee the security of a blockchain system, we evaluate the robustness of RBBC by implementing Byzantine attacks and assessing empirically the behavior of RBBC. In a distributed system, the misbehavior of some machines, that could be due to a simple misconfiguration, may affect the result of the entire computation, like the consensus decision. To observe the robustness of RBBC, we implemented alternative Red Belly Blockchain programs flipping the bits that they should send to slow down the consensus execution and sending wrong information to delay the broadcast of the information among correct nodes. We deployed these intentionally misbehaving codes and observed the Byzantine fault tolerance of RBBC and the impact on performance these misbehaviors could cause.

Finally, we compared the RBBC against a classic leader-based BFT~\cite{CL02,BSA14,Kwo14} protocol and HoneyBadgerBFT~\cite{MSC16}. The results indicate a latency among the lowest but significantly higher throughput.  More specifically, 
in the same settings the throughput of RBBC peaks at hundreds of machines where the traditional Byzantine fault tolerant solutions would not scale to 40 machines and the latency of RBBC is a third of the HoneyBadgerBFT randomized alternative.

We start with the background on other blockchain systems (\cref{sec:rw}) and present the design decision of the Red Belly Blockchain (\cref{sec:design}).
We then describe the settings of our experiments (\cref{sec:evaluation}).
We compare the Red Belly Blockchain to both leader-based and randomized blockchains when deployed on geo-distributed machines and illustrate the performance gained from its transaction signature verification and its proposal combination to decide large blocks (\cref{sec:geo1}).
We then evaluate the security of the Red Belly Blockchain by running Byzantine attacks and observing its resilience and how it maintains performance (\cref{sec:security}).
Finally, we present the performance on a thousand of virtual machines (\cref{ssec:thousand-vm}) and we conclude (\cref{sec:conclusion}).

\section{Background}\label{sec:rw}

Byzantine Fault Tolerance (BFT) requires a number of messages per consensus instance that grows quadratically with the number of nodes. 
Most previous work on Byzantine-fault tolerant blockchains~\cite{Kwo14,BSA14,SBV17} would solve the traditional 
Byzantine consensus problem~\cite{LSP82}, deciding only one proposed value, regardless of the number of participants~\cite{CL02,BSA14,MA06,ABQ13,AGK14}.
Due to this scalability limitation~\cite{Vuc15,Buc16,SBV17}, we see various blockchain proposals reverting to a crash tolerant model to tolerate more but simpler failures, like Hyperledger Fabric~\cite{ABB18}. Not tolerating Byzantine failures confines these blockchains to 
secure networks that are protected from intrusions by other means.

For the sake of scalability, one can instead solve a variant of the consensus problem that allows to 
combine proposals into a decision~\cite{BKR94,NCV05,CGLR18}. 
Extra care is however needed before one can apply solutions to these problems to decide a ``super'' block combining multiple proposed blocks. 
For example, the related problems of Agreement on a Core Set or Asynchronous Common Subset (ACS)~\cite{BKR94}, Interactive Consistency (IC)~\cite{LSP82} and Vector Consensus (VC)~\cite{NCV05} require either $n-t$ or  $t+1$ (at least $t+1$ as $n>3t$) proposed values to be decided. In blockchain, however, there may not even be $t+1$ compatible proposed blocks. 
This incompatibility arises
when at least one transaction per block is not correctly signed or transactions of two distinct blocks conflict. Accepting to commit these invalid transactions could limit fairness and introduce starvation,
yet at proposal time even correct proposers cannot anticipate these conflicts.
Instead of trying to decide a minimal number of proposed blocks,
RBBC propose sets of transactions and decides on a new block whose transactions are the union of correctly signed and non-conflicting proposed transactions.

\sloppy{Many blockchains~\cite{Nak08,Woo15,LNZ16,AMN17,KJG17,GHM17} assume synchrony. The drawback is that 
if the messages experience an unforeseen delay, then the blockchain guarantees are violated.
These delays can for example be exploited in proof-of-work blockchains~\cite{Nak08,Woo15} to double spend~\cite{NG17}.
To avoid incentivizing all nodes to generate a proof-of-work that waste CPU resources, 
alternative proof-of-* models were proposed.
Algorand~\cite{GHM17} uses randomization to restrict the task of deciding a block to a small subset.
Elastico~\cite{LNZ16} proposes a sharded consensus partitioned into sub-committees to run fast but more consensus instances.
By contrast, the RBBC sharding only applies to verifiers and proposers of a single instance, which gives a low latency and a high throughput.}

Recent blockchains try to avoid the synchrony assumptions~\cite{Kwo14,MSC16,SBV17,ABB18}.
The HoneyBadger Byzantine Fault Tolerance (HBBFT)~\cite{MSC16} 
aims at solving ASC by building upon an asynchronous binary Byzantine consensus algorithm~\cite{MMR14} that is probabilistic 
and assumes a fair scheduler~\cite{MMR15}. 
As we show in \cref{sec:geo1}, even with a fair scheduler HBBFT is too costly for RBBC to build upon it because it requires a binary consensus~\cite{MMR14} that requires a common coin.
Some alternatives relax this fair scheduler assumption but they require more messages, which risks to increase  the overhead~\cite{MMR15}.
To avoid both randomization and synchrony, various solutions~\cite{CL02,KADC07,CMS07,VCB09,MJM09,AGK14,BSA14} assume partial 
synchrony~\cite{DLS88}. 
Unfortunately, they
all rely on a leader or a primary to propose to others and follow a three-step execution pattern that offers low latency but whose throughput cannot scale. The lack of scalability of this pattern was experimented in~\cref{sec:evaluation}.

Recent blockchains, like Tendermint~\cite{Kwo14} achieve security by building upon these BFT protocols.
As they all inherit the same execution pattern, they all decide only one of the proposed set of transactions and do not scale to tens of nodes~\cite{Vuc15,Buc16,SBV17}. 
A Byzantine fault tolerant ordering service was tested for Hyperledger Fabric, however, its best performance was achieved with $n=4$ nodes~\cite{SBV17}.
Hyperledger Fabric finally reverted to a crash fault tolerance version~\cite{ABB18} to 
try to scale to a larger network. Despite the lack of Byzantine fault tolerance, its consensus was only evaluated across two data centres~\cite{ABB18}.
By contrast, RBBC features a Democratic BFT (DBFT) that extends a Byzantine consensus algorithm~\cite{CGLR18} by creating a super block resulting from multiple proposed sets of transactions.
The reason for choosing this consensus algorithm is that it does not follow the common leader-based pattern
that has costly recoveries in the case of faulty or slow leaders~\cite{CWA09,BS09,ABQ13}.

Despite these differences, RBBC combines many of the optimizations proposed in the 
aforementioned BFT literature.
Its leaderless DBFT algorithm stems from a provably correct algorithm~\cite{CGLR18}, its concurrent implementation leverages multicores similar to BFT-Smart~\cite{BSA14}. It generalises the $n$ proposers of HoneyBadger BFT to any number~\cite{MSC16}.
In addition, RBBC could potentially benefit from other BFT optimizations.
Some proposals~\cite{SBV17} allow temporary inconsistencies and transaction rollback. 
Other proposals rely on trusted components~\cite{VCB10,MJM09,BDK17}.
In particular, RAM~\cite{MJM09} suggests to use the Attested Append-Only Memory (A2M)~\cite{CMS07} trusted service to scale, however, we are not aware of any implementation.
Steward~\cite{ADD06} organizes consensus into a hierarchy to scale to wide area networks and offers tens of updates per second on an emulated wide area network. 
RBBC extends this related work through the implementation of a replicated state machine tailored for blockchain.

\section{The Design of the Red Belly Blockchain}\label{sec:design}

The Red Belly Blockchain was initially presented at MIT in July 2017 where a first version of the system could achieve 440,000 transactions per second on 100 machines. The performance was optimized and shown to achieve 660,000 transactions per second on 300 machines at Facebook and Visa Research in October 2017 as we explain in this paper. The details of these presentations are available online~\cite{Gra17}.

In short, the Red Belly Blockchain is a community blockchain~\cite{VG18} with a dynamic set of consensus participants or proposers whose public keys are listed in a configuration block. These proposers receive from permissionless clients some balance, subscription and transaction requests.  Proposers can answer balance requests based on the information they have about the current state of the blockchain and they keep a list of subcribers to send them updates about the balance of all accounts. Both proposers and subscribers are called \emph{replicas} as they maintain a copy of the state of the blockchain either as the full blockchain or as a UTXO table. Proposers store transactions in a memory pool or \emph{mempool} before proposing them to some consensus instance. Once a client receives an identical balance response from $t+1$ proposers, it knows the balance of its account. Once the consensus decides upon a combination of the proposed transactions that are correctly signed and not in conflict, this combination is wrapped into a block appended to the chain.

\subsection{Optimized Democratic BFT}\label{ssec:dbft}
The unprecedented performance of the Red Belly Blockchain is mainly due to a novel design that relies on a Byzantine consensus algorithm especially designed for blockchains and called \emph{Democratic Byzantine Fault Tolerance (DBFT)}~\cite{CGLR18} that does not assume synchrony. Most Byzantine consensus algorithms predate the blockchain era and were not designed to scale to a very large number of machines as it is needed in blockchains. 

For this reason, there are few distinctions between DBFT and classic Byzantine fault tolerant algorithms.

First, DBFT does not rely on a leader to avoid any bottleneck effect at large scale. Instead it allows multiple proposers to propose disjoint sets of transactions that could all be inserted in the block decided at the end of the consensus.

Second, DBFT solves the consensus deterministically. Hence the Red Belly Blockchain never \emph{forks}, a situation where multiple blocks are appended at the same index of the chain and that could be exploited by attackers to double-spend~\cite{NG17}.  In particular, it does neither require a common coin nor a fair scheduler. The interested reader can access the detailed proofs in the technical report~\cite{CGLR17}.

To reduce the bandwidth usage of the reliable broadcast of DBFT~\cite{CGLR18}, we included a SHA256 hash digest of the message 
instead of including the full proposal in the $\lit{echo}$ and
$\lit{ready}$ messages. To increase throughput we grouped all valid and non-conflicting transactions obtained at the end of the consensus to create a new block.

\subsection{Sharded Verification}
The Red Belly Blockchain also offers other advantages as it shards the verification of transaction signatures.  Traditional blockchains either require all active participants to verify the signature of each individual transaction or assume the presence of trusted verifiers or endorsers. The Red Belly Blockchain leverages the computational resources of the participants by spreading the load of verifying transactions to different subsets of participants but without requiring trust. It only requires each transaction to be verified by at least $t+1$ participants but never more than $2t+1$ participants.

The Red Belly Blockchain is a full-fledge blockchain that supports UTXO transactions signed through Elliptic Curve Digital Signature Algorithm (ECDSA) and verified at run-time. All communications are encrypted through SSL, which does not impact performance significantly. It offers a model of open permissioned blockchain called \emph{community blockchain} in that it relies on a dynamic set of participants whose public key are well identified to run the consensus but allows permissionless clients to issue transaction and balance requests. More details on how this community blockchain bypasses the predetermined set of participants requirement of consortium blockchain can be found in~\cite{VG18}.

The optimized deterministic leader-less DBFT consensus designed for blockchain and the sharded verification allows Red Belly Blockchain to be a secure blockchain that does not fork and whose performance scales with the amount of computational resources coming with hundreds of participants.

\section{Experimental Settings}\label{sec:evaluation}
In this section, we evaluate RBBC on up to 1000 machines on Amazon EC2 located in up to 
14 separate regions.
To this end, we compare the performance of (1)~RBBC with its sharding and its DBFT consensus.
(2)~RBBC where we replaced DBFT by the Honey 
Badger of BFT 
protocol (HBBFT)~\cite{MSC16}, for which we reused the publicly available cryptographic operations 
implementation and
(3)~RBBC where we replaced DBFT by a classic 3-step leader-based BFT algorithm \textit{CONS1}~\cite{CL02,Kwo14,BSA14}. 

We run three types of experiments: \textit{(i)}~with up to 300 deciders all deciding and generating the workload, allowing
new proposals to be made as soon as the previous one is committed (\cref{sec:geo1} and~\cref{sec:datacenter}); 
\textit{(ii)}~with requesters running on nodes separated from the permissioned nodes to measure their impact on performance and finally (\cref{ssec:remote-req});
\textit{(iii)}~with up to 1000 nodes all runnings as replicas, some requesting and some deciding, but all updating their copy of all account balances (\cref{ssec:thousand-vm}).

\subsection{Leader-based and randomized BFT}
CONS1 is the classic 3-step leader-based Byzantine consensus implementation similar to PBFT~\cite{CL02}, the Tendermint consensus~\cite{Kwo14}, and including the concurrency optimizations of BFT-Smart~\cite{BSA14}. 
To reduce network consumption CONS1 is implemented using digests in messages that follow the initial broadcast.

The HoneyBadger Byzantine Fault Tolerance (HBBFT)~\cite{MSC16} 
aims at solving the ASC problem~\cite{BKR94} by building upon an asynchronous binary Byzantine consensus algorithm~\cite{MMR14} that is probabilistic 
and assumes a fair scheduler~\cite{MMR15}.
To evaluate HBBFT we used the source code provided by the authors of HBBFT.

Both CONS1 and HBBFT variants make use of a classic verification, as in traditional blockchain 
systems~\cite{Nak08,Woo15},
that takes place at every decider upon delivery of the decided block from consensus.
Unless otherwise stated, all nodes behave correctly.
Apart from the sharded verification of RBBC, all algorithms
run the same code for the state-machine component implementing the blockchain.
Note that there exist BFT algorithms that terminate in less message steps
than CONS1, but require additional assumptions like non-faulty clients~\cite{AGK14,KADC07}
or $t<n/5$~\cite{MA06}.
HBBFT 
uses
a randomized consensus~\cite{MMR14} and reliable broadcast using erasure codes.

\begin{table*}
  \tiny{
\setlength{\tabcolsep}{6pt}
\newcommand\items{15}   
\arrayrulecolor{white} 
\noindent\begin{tabular}{c*{\items}{|E}|}
\multicolumn{1}{c}{} & 
\multicolumn{1}{c}{Tokyo} & 
\multicolumn{1}{c}{Seoul} & 
\multicolumn{1}{c}{Mumbai} & 
\multicolumn{1}{c}{Singa.} & 
\multicolumn{1}{c}{Sydney} &
\multicolumn{1}{c}{Canada} &
\multicolumn{1}{c}{Frankfurt} & 
\multicolumn{1}{c}{Ireland} & 
\multicolumn{1}{c}{London} &
\multicolumn{1}{c}{S\~ao P.} &
\multicolumn{1}{c}{N.Virg.} & 
\multicolumn{1}{c}{Ohio} &
\multicolumn{1}{c}{N.Cal.} &
\multicolumn{1}{c}{Oregon} 
\\ \hhline{~*\items{|-}|}
Tokyo & 0   & 551 & 129 & 240 & 161 & 106 & 74 & 66.4 & 59 & 55.4 & 90.1 & 96.2 & 129 & 132 \\ \hhline{~*\items{|-}|}
Seoul  & 33   & 0  & 137 & 157 & 141 & 91.5 & 54 & 60.8 & 54.7 & 56.6 & 84.2 & 114 & 84.2 & 116 \\ \hhline{~*\items{|-}|}
Mumbai  & 133   & 164  & 0 & 121 & 67 & 90.9 & 176 & 178 & 145 & 46.7 & 81.9 & 80.5 & 69.1 & 64.2 \\ \hhline{~*\items{|-}|}
Singapore  & 69   & 100  & 67 & 0 & 90.9 & 83.2 & 90.7 & 86.1 & 90.4 & 40.8 & 59.5 & 64.9 & 80.5 & 77.3 \\ \hhline{~*\items{|-}|}
Sydney  & 106  &  135  &  235  &  170  &  0 & 77.1 & 61.3 & 53.8 & 51.2 & 40.2 & 74.9 & 99.7 & 135 & 119 \\ \hhline{~*\items{|-}|}
Canada &  166  &  185  &  196  &  220  &  225  &  0 & 166 & 250 & 164 & 159 & 808 & 760 & 205 & 168 \\ \hhline{~*\items{|-}|}
Frankfurt &  244  &  275  &  112  &  178  &  292  &  102  &  0 & 477 & 823 & 92.9 & 222 & 220 & 144 & 85.7\\ \hhline{~*\items{|-}|}
Ireland &  226  &  246  &  122  &  188  &  286  &  78  &  25  &  0 & 829 & 114 & 185 & 183 & 104 & 117\\ \hhline{~*\items{|-}|}
London &  255  &  284  &  111  &  179  &  281  &  90  &  15  &  12  & 0  & 107 & 190 & 195 & 107 & 85.5\\ \hhline{~*\items{|-}|}
S\~ao Paulo &  271  &  293  &  302  &  328  &  332  &  125  &  210  &  184  &  192  &  0 & 131 & 124 & 77.7 & 81.7\\ \hhline{~*\items{|-}|}
N. Virginia  &  162  &  209  &  182  &  238  &  205  &  15  &  89  &  85  &  76  &  122  & 0  & 827 & 232 & 186\\ \hhline{~*\items{|-}|}
Ohio &  169  &  199  &  193  &  227  &  196  &  25  &  99  &  91  &  87  &  131  &  13  & 0  & 428 & 219\\ \hhline{~*\items{|-}|}
N. California &  120  &  150  &  262  &  178  &  148  &  76  &  148  &  142  &  138  &  182  &  64  &  52  &  0 & 681\\ \hhline{~*\items{|-}|}
Oregon &  105  &  135  &  235  &  163  &  162  &  66  &  164  &  141  &  158  &  183  &  76  &  71  &  22  &  0\\ \hhline{~*\items{|-}|}
\end{tabular}}
    \caption{Heatmap of the bandwidth (Mbps) in the top right triangle and latency (ms) in the bottom left triangle between the 14 regions of Amazon Web Services, as used in our experiments\label{table:pings}}
   \vspace{-2em}
\end{table*}

\subsection{Machine specification}
We run the blockchains on the 14 Amazon datacenters that we had at our disposal at the time of the experiment: North Virginia, Ohio, North California, Oregon, Canada, Ireland, Frankfurt, London, Tokyo, Seoul, Singapore, Sydney, Mumbai, S\~ao Paulo.
We tested two different VMs: (1)~\emph{high-end} c4.8xlarge instances with an Intel Xeon E5-2666 v3 processor of 18 hyperthreaded cores,
60 GiB RAM and 10\,Gbps network performance when run in the same datacenter where storage is backed by Amazon's Elastic Block Store (EBS) with 4 Gbps dedicated throughput;
(2)~\emph{low-end} c4.xlarge instances with an Intel Xeon E5-2666 v3 processor of 4 vCPUs, 7.5 GiB RAM, and ``moderate'' network performance (as defined by Amazon). Storage is backed by EBS with 750\,Mbps dedicated throughput.
To limit the bottleneck effect on the leader of PBFT, we always place the leader in the most central (w.r.t. latency) region, Oregon. When not specified, proposals contain 10,000 transactions and $t$ is set to the larger integer strictly lower than $\frac{n}{3}$.

\section{Comparing geodistributed blockchains}\label{sec:geo1}

First, we report the performance when running 10 high-end VMs in each of the 14 regions for a total of 140 machines.
%
At the time of this experiment Amazon was offering us only 14 availability zones: North Virginia, Ohio, North California, Oregon, Canada, Ireland, Frankfurt, London, Tokyo, Seoul, Singapore, Sydney, Mumbai, S\~ao Paulo.
Each zone contains 10 high-end machines. 
As depicted on Table~\ref{table:pings}, we computed the variation of communication latencies and throughput between these Amazon EC2 datacenters as measured using c4.xlarge instances.
The minimum latency is $11$\,ms between London and Ireland, whereas the maximum latency is $332$\,ms observed between Sydney and S\~ao Paulo.
Bandwidth between Ohio and Singapore is measured at approximately 64.9 Mbits/s (with variance between 6.5 Mbits/s and 20.4 Mbits/s).

\begin{figure}[t]
        \begin{center}
        \includegraphics[scale=0.65]{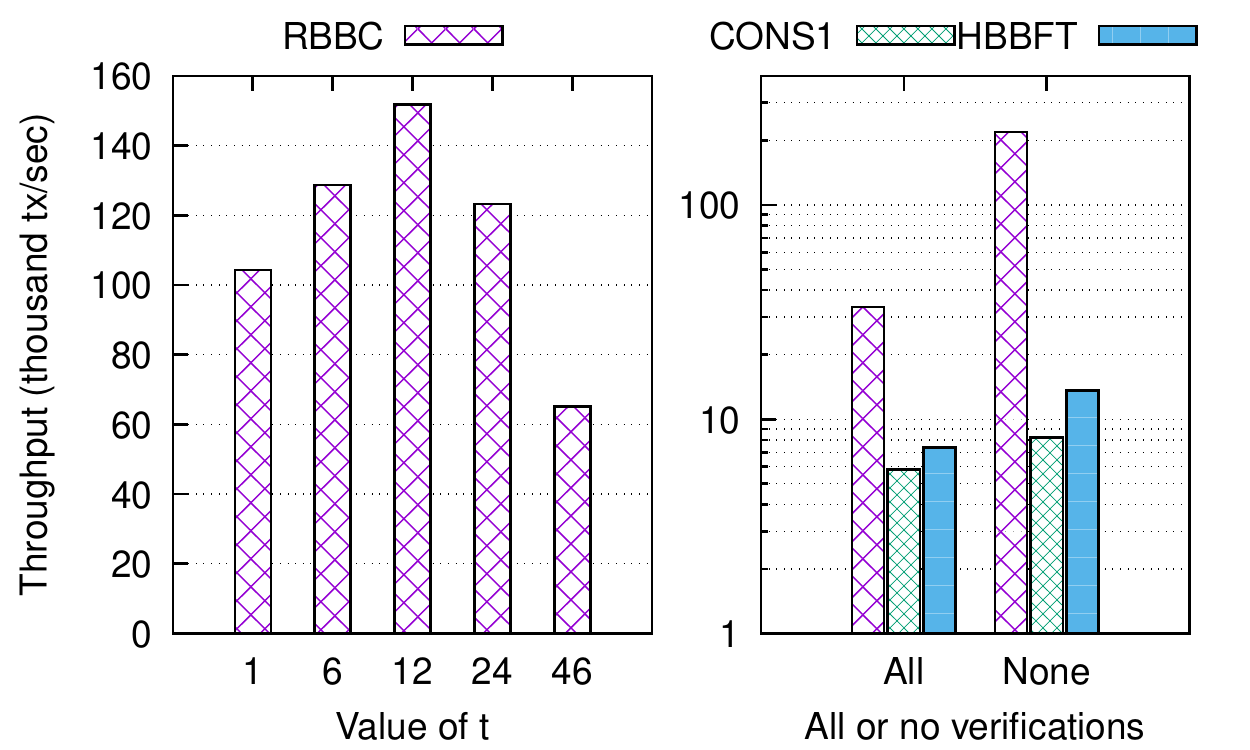}
        \caption{Impact of fault tolerance and verification on the RBBC throughput when $n=140$ geodistributed machines 
        \label{fig:consensus-fault-tolerance}}
        \end{center}
\end{figure}

\subsection{Impact of verification}\label{ssec:verification-cost}
To measure the impact of verification on performance, we varied 
the parameter $t$ from the minimum to its maximum value ($46 < \frac{140}{3}$) 
with sharded verification as depicted in Figure~\ref{fig:consensus-fault-tolerance} (left)
and we compared all three blockchains with all nodes verifying all transactions (all) and with 
no verification (no validation) as depicted in Figure~\ref{fig:consensus-fault-tolerance} (right).
The peak throughput of 151,000 \emph{transactions per second} (tx/sec) is achieved with the fault-tolerance parameter $t=12$. 
When $t\leq 6$, performance is limited by the $(t-1)^{th}$ slowest node as the consensus
waits for a higher number of $n-t$ proposers.
When $t\geq 24$, performance is then limited by the growing number of $t+1$ necessary verifications. 
In Figure~\ref{fig:consensus-fault-tolerance} (right), the performance of all algorithms is higher 
without verification than with full verification. RBBC is the most affected dropping from 
219,000\,tx/sec to 33,000\,tx/sec while HBBFT and CONS1 throughputs drop less but from a lower 
peak. As we will 
show in~\cref{sec:proposals-gain} and \cref{sec:geo2},
there are factors other than verification that have a larger impact on these algorithms.

\begin{figure}[t]
        \begin{center}
          \includegraphics[scale=0.65]{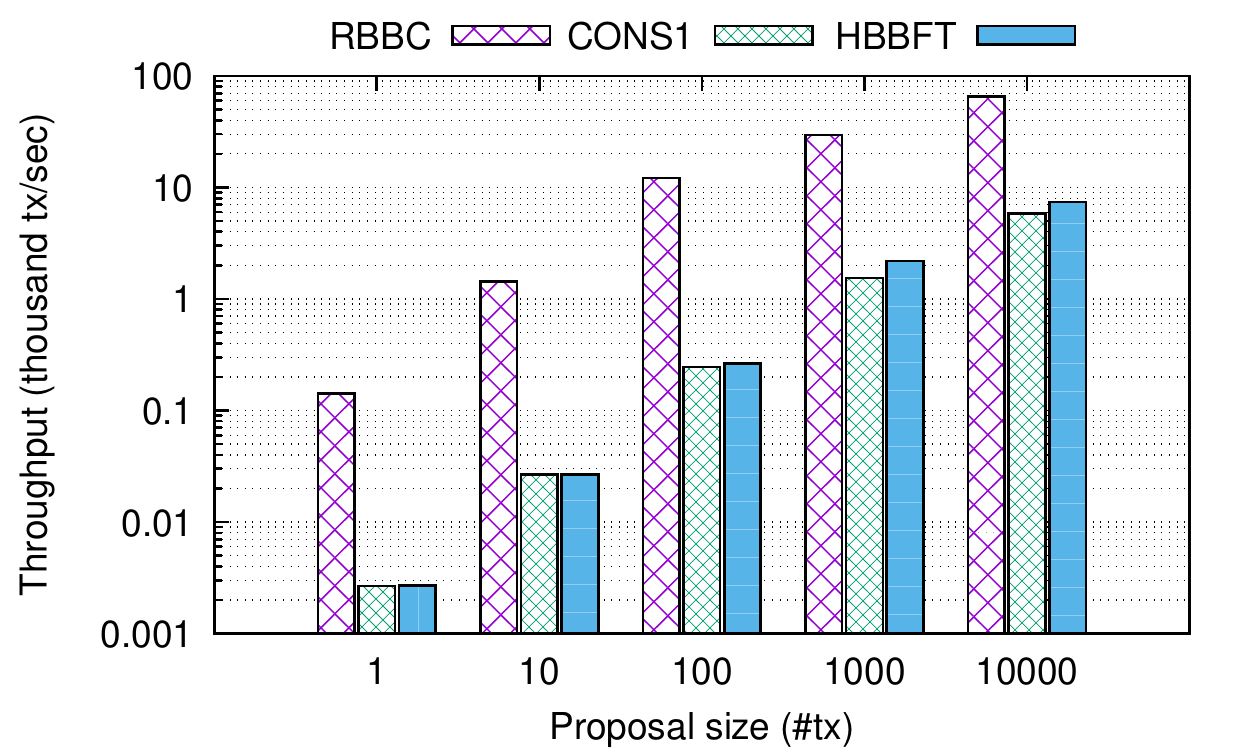}
          \includegraphics[scale=0.65]{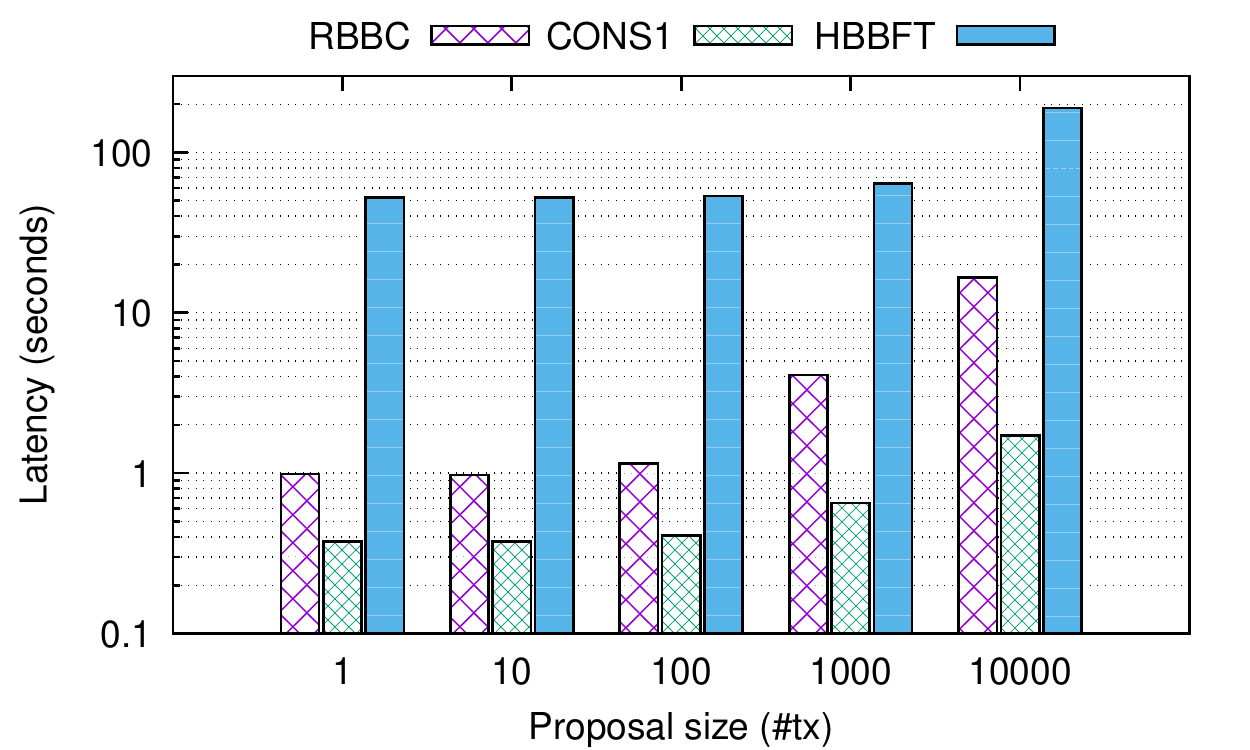}          
          \caption{Throughput and latency comparison of the blockchain solutions with $n=140$ and $t=46$, and proposal sizes of $1,10,100,1000$ and $10000$\label{fig:consensus-comparison-valid}}
        \end{center}
\end{figure}

\subsection{Combining proposals}\label{sec:proposals-gain}

Figure~\ref{fig:consensus-comparison-valid} explored the effect of deciding the unions of proposals when running the blockchain.
CONS1 has the lowest latency because in all executions the leader acts
correctly, allowing it to terminate in only 3 message delays, where RBBC with DBFT requires 4 message delays.
Probably due to its inherent concurrency, RBBC offers the best latency/throughput tradeoff: at 1000\,ms latency, RBBC offers 12,100\,tx/sec whereas at 1750\,ms latency, CONS1 offers only 5800\,tx/sec.
Finally, the blockchain with HBBFT has the worst performance for several reasons: HBBFT relies on a consensus algorithm~\cite{MMR14} whose termination is randomized and it uses erasure codes:
the computation time needed for performing reliable broadcast using erasure codes on a single node with a proposal size of 1000 transactions takes over 200\,ms.  Each node then has to do this for each proposal (i.e., 140 times in this experiment) increasing significantly the latency.  

\begin{figure}[t]
        \begin{center}
          \includegraphics[scale=0.65]{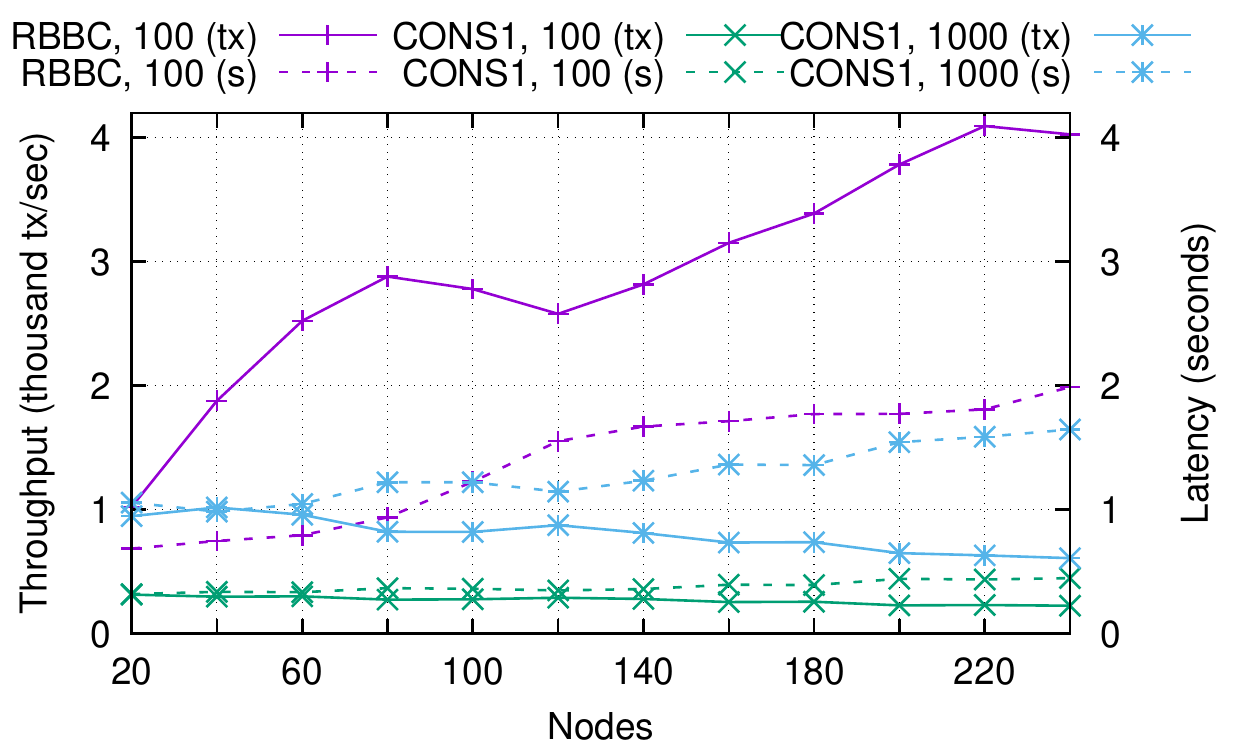} \label{fig:t+1-100sca}
          \includegraphics[scale=0.65]{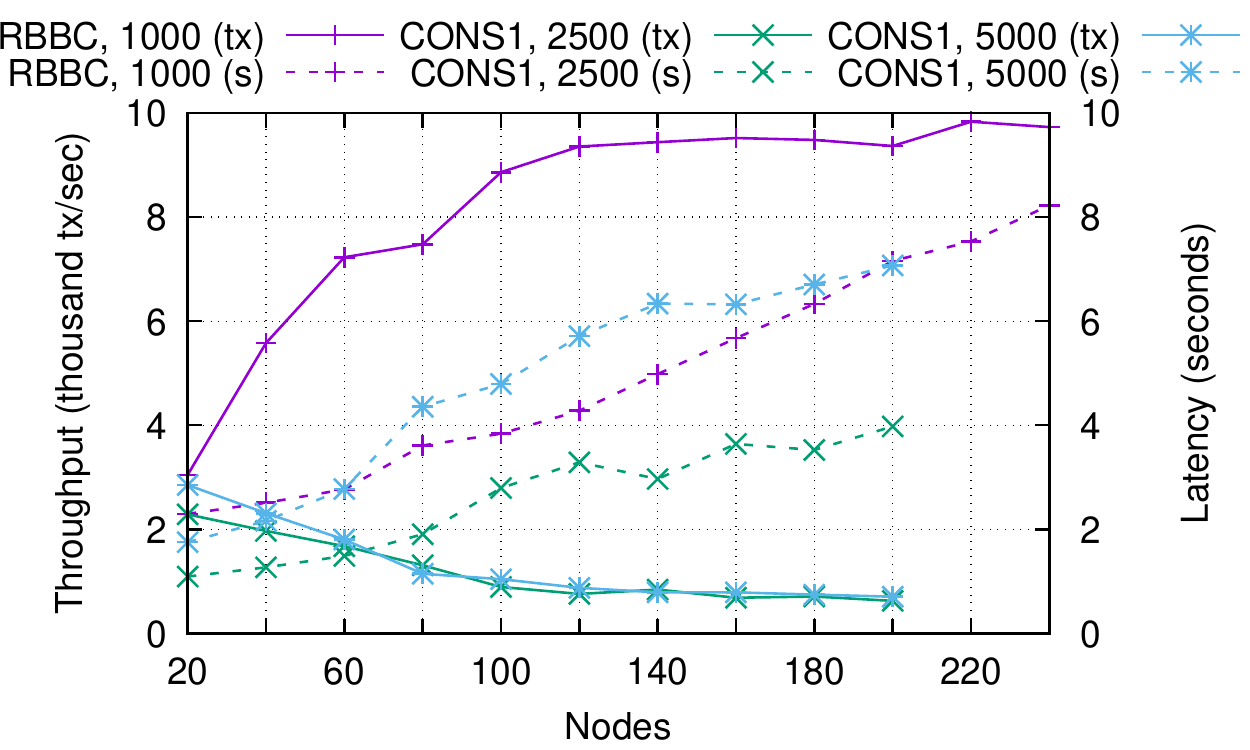} \label{fig:t+1-1000sca}
          \caption{The performance of CONS1 and RBBC with $t+1$ proposer nodes; the number following the algorithm name represents the number of transactions in the proposals;
            solid lines represent throughput, dashed lines represent latency\label{fig:t+1scalability}}
        \end{center}
   \vspace{-1.5em}
\end{figure}

\subsection{Low-end machines and distributed proposals}\label{sec:geo2}

We now experiment on up to 240 low-end VMs, whose CPU resource is closer to the one of cell phones, and evenly spread on 5 datacenters in the United States (Oregon, Northern California, and Ohio) and Europe (Ireland and Frankfurt).
We examine the impact of having $t+1$ vs. $n$ proposers.
Dedicating the 4 vCPUs of these low-end instances led to verify about 7800 serialized transactions per second with 97\% of CPU time spent verifying signatures and with 3\% spent
deserializing and updating the UTXO table.  

\begin{figure}[t]
        \begin{center}
          \includegraphics[scale=0.65]{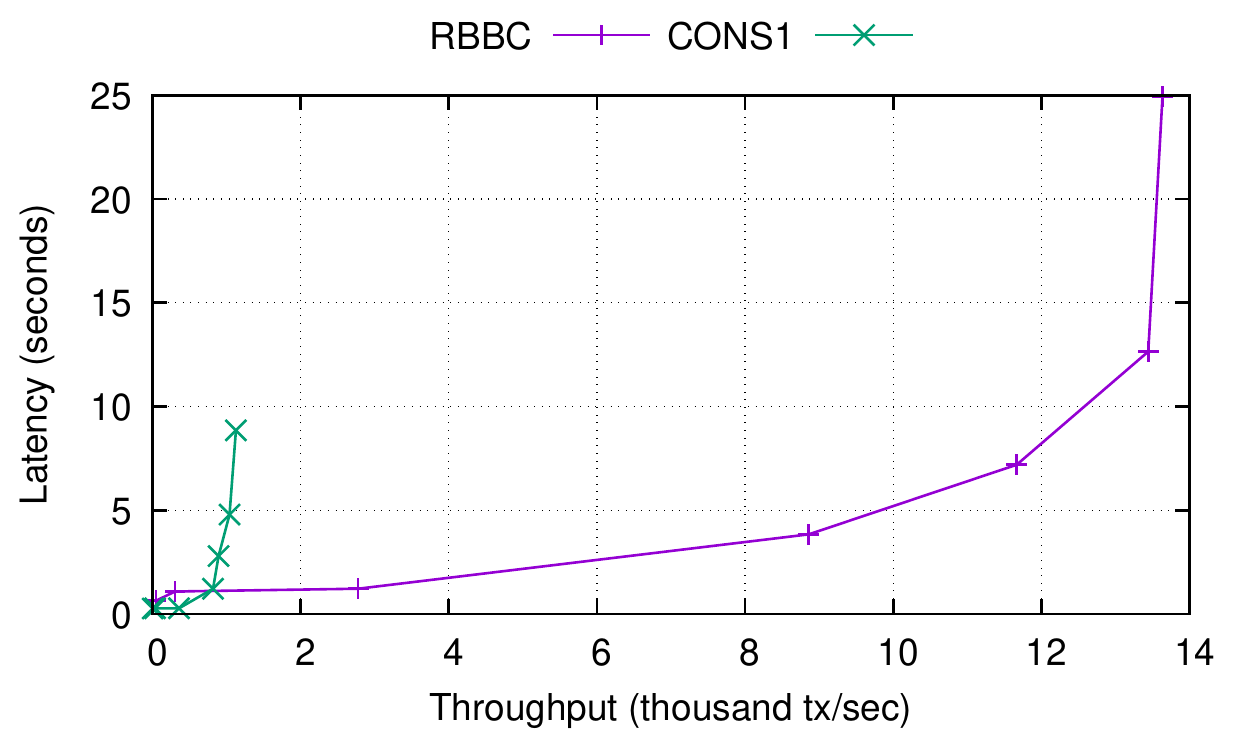}
          \caption{Comparing throughput and latency of CONS1 and RBBC with $t+1$ proposer nodes on 100 geodistributed nodes;
            each point represents the number of transactions in the proposals, either 10, 100, 1000, 2500, 5000 or 10000 (HBBFT does not appear due to lower performance)\label{fig:lat-tp-100}}
        \end{center}
   \vspace{-1.5em}
\end{figure}

\subsection{The impact of $t+1$ proposer nodes}
Figure~\ref{fig:t+1scalability} shows the throughput and latency 
of RBBC with $t+1$ proposers and CONS1 with different sizes of proposals.
As CONS1 is limited to a single proposer (its leader) while RBBC supports multiple proposers, 
we tested whether CONS1 performance would be better with more transactions per proposal 
than RBBC.     

With proposal size of 100, RBBC throughput increases from 1000 to 4000\,tx/sec while its latency increases from 750\,ms to 2\,seconds.
The throughput increase stems from increasing CPU and bandwidth resources with more proposers.
With larger proposal size (1000), performance increases faster (from 3000\,tx/sec to 9000\,tx/sec) with the number of nodes and flattens out earlier around 10,000\,tx/sec while 
latency increases from 2 to 8 seconds.

With proposal size of 100, CONS1 throughput decreases from 310\,tx/sec to 220\,tx/sec while latency increases from 320\,ms to 460\,ms.
Unfortunately, this low latency does not help to increase throughput by increasing proposal size after a certain number of nodes.
In particular, with proposal size of 5000 the throughput drops by 4 times (from 2800\,tx/sec to 700\,tx/sec). While CONS1 can broadcast message authentication codes (MACs) through UDP in local area networks, no such broadcast primitive is available in this wide area testnet.

Figure~\ref{fig:lat-tp-100} further examines the performance of CONS1 and RBBC with 100 nodes and proposal sizes of 1, 10, 100, 1000, 2500, and 5000.
Here we see that the throughput of CONS1 reaches a limit of about 1100\,tx/sec while RBBC approaches 14,000\,tx/sec.  CONS1 has a better minimum latency of
270\,ms compared to 640\,ms for RBBC for proposals of size $1$.

\begin{figure}[t]
        \begin{center}
          \includegraphics[scale=0.65]{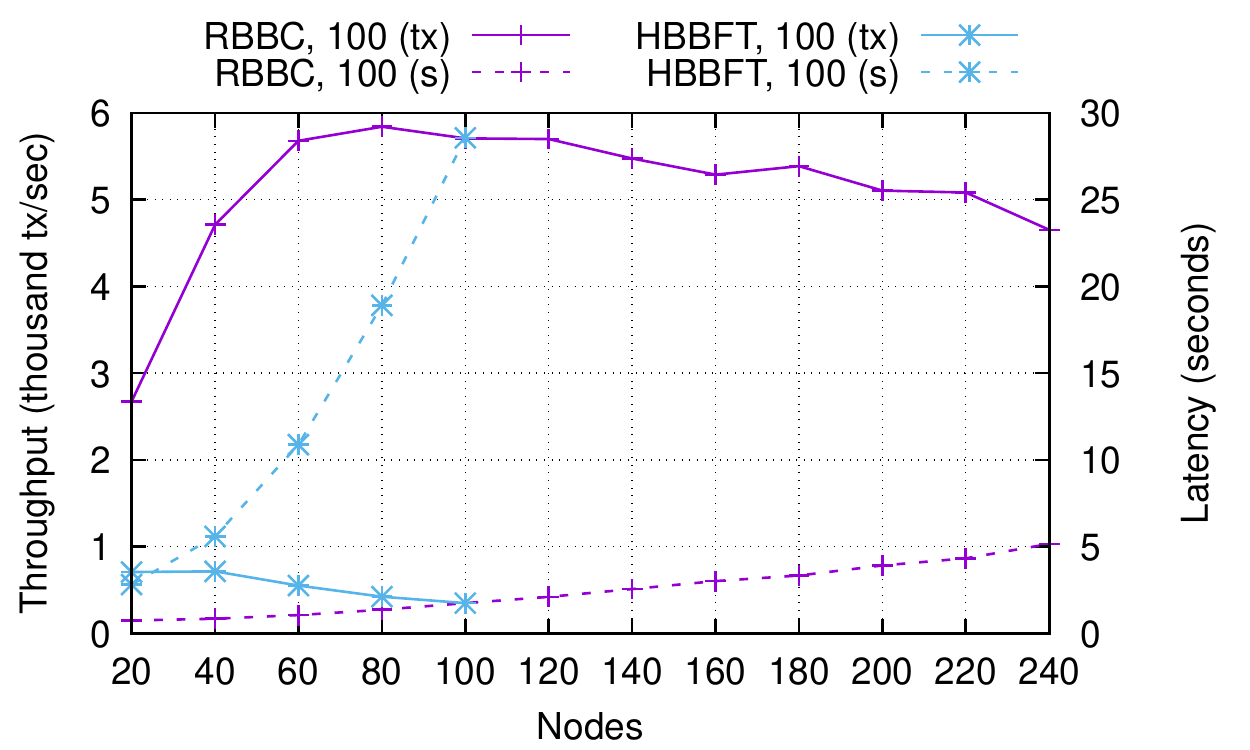} \label{fig:n-100sca}
          \includegraphics[scale=0.65]{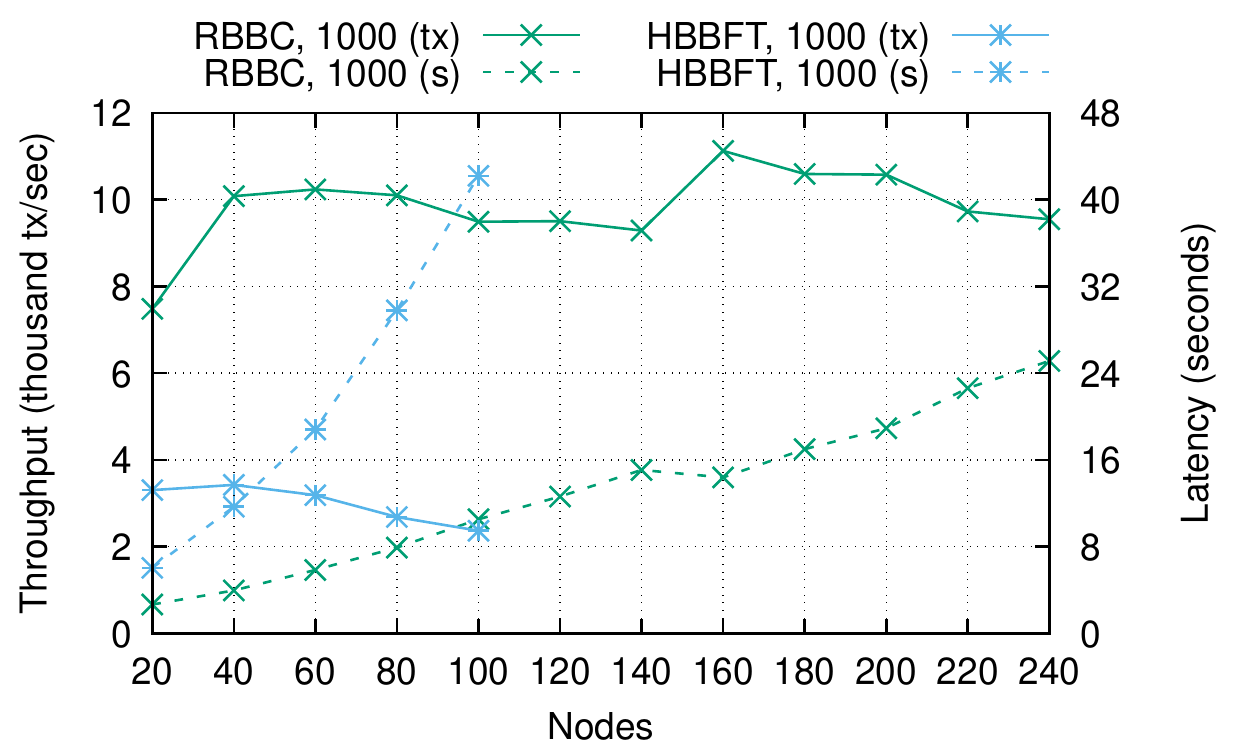} \label{fig:n-1000sca}
          \caption{The performance of HBBFT and RBBC with $n$ proposer nodes.  The number following the algorithm name represents the number of transactions in the proposals;
            solid lines represent throughput, dashed lines represent latency \label{fig:nscalability}}
        \end{center}
\end{figure}

\subsection{The impact of $n$ proposer nodes}
Figure~\ref{fig:nscalability} depicts the performance 
of RBBC and HBBFT with $n$ proposers,
with proposal sizes of 100 and 1000 transactions.
Unsurprisingly, with $n$ proposers the throughput of RBBC increases faster than with $t+1$ proposers.
With a proposal size of 100, the throughput reaches 6000\,tx/sec at 80 nodes and slowly degrades, while latency starts at 740\,ms with 20 nodes and reaches 5160\,ms with 240 
nodes.
With a proposal size of 1000, the throughput reaches 10,000\,tx/sec at 40 nodes and remains mostly flat, latency starts at 2670\,ms with 20 nodes and reaches 25,100\,ms with 240 nodes.
With larger node counts (around 200), the configurations with $t+1$ proposals achieve similar throughput, but with much lower latency.
Thus when using nodes similar to the low-end instances, having $n$ proposers seems better suited for configurations of less than 100 nodes.

For HBBFT we observe that latencies increase superlinearly and throughput degrades as we increase the number of nodes. As mentioned before, this is primarily due to the computation
needed for the erasure codes.
Note that we only run HBBFT up to 100 nodes as afterwards we start seeing latencies approaching minutes.

\subsection{Transaction verification count}\label{ssec:verification-count}
\begin{figure}[t]
        \begin{center}
          \includegraphics[scale=0.65]{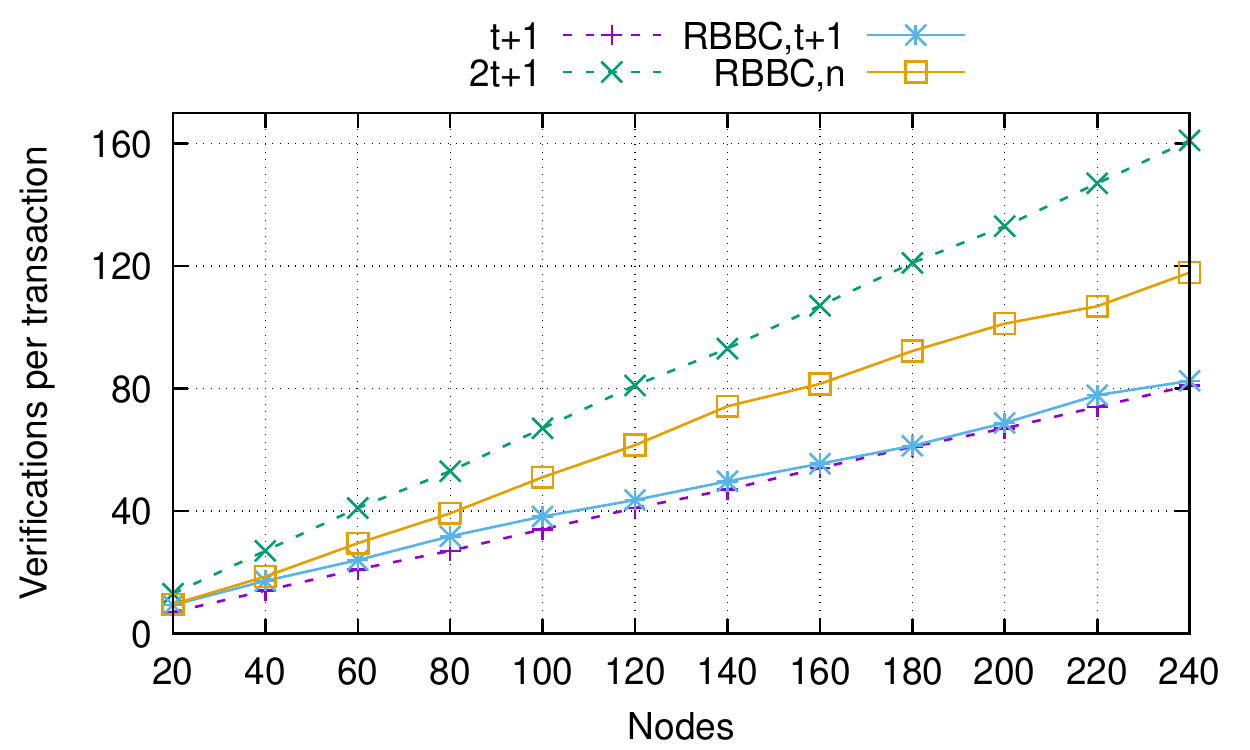}
          \caption{The number of times a transaction is verified in RBBC with proposal size of 100 transactions, with either $t+1$ or $n$ proposer nodes; the dashed lines $t+1$ and $2t+1$ represent the minimum and maximum number of possible verifications.
            \label{fig:sca-val}}
        \end{center}
   \vspace{-2em}
\end{figure}

In the previous experiments we also recorded the average number of times a transaction is verified to examine the state
of sharded verification, the results are shown in Figure~\ref{fig:sca-val}.
The best case is $t+1$ verifications while the $2t+1$ is the worst case.
We observe that with $t+1$ proposers the number of verifications stays close to the optimal, while with $n$
proposers the number of verifications remains around the middle of $t+1$ and $2t+1$.
This is likely due to the increased load on the system causing verifications to occur in different orders at different nodes.

\section{Experiment under Byzantine attacks}\label{sec:security}

We evaluate RBBC performance under 2 Byzantine attacks:
\begin{enumerate}
\item [{\bf Byz1}] The payload of the reliable broadcast messages altered so that no proposal is delivered for reliable broadcast instances led by faulty nodes.
  The binary payloads of the binary consensus messages are flipped. The goal of this behavior is to reduce throughput and increase latency.
\item [{\bf Byz2}] The Byzantine nodes form a coalition in order to maximize the bandwidth cost of the reliable broadcast using the digests described in \cref{ssec:dbft}. 
As a result, for any reliable broadcast initiated by a Byzantine node, $t+1$ correct nodes will deliver the full
message while the remaining $t$ will only deliver the digest of the message, meaning they will 
have to request the full message from $t+1$ different nodes from whom they receive $\lit{echo}$ messages.
\end{enumerate}
As in \cref{sec:geo2}, experiments are run with 100 low-end machines using the same 5 datacenters from US and Europe and with $n$ proposers.

\begin{figure}[t]
       \begin{center}
         \includegraphics[scale=0.65]{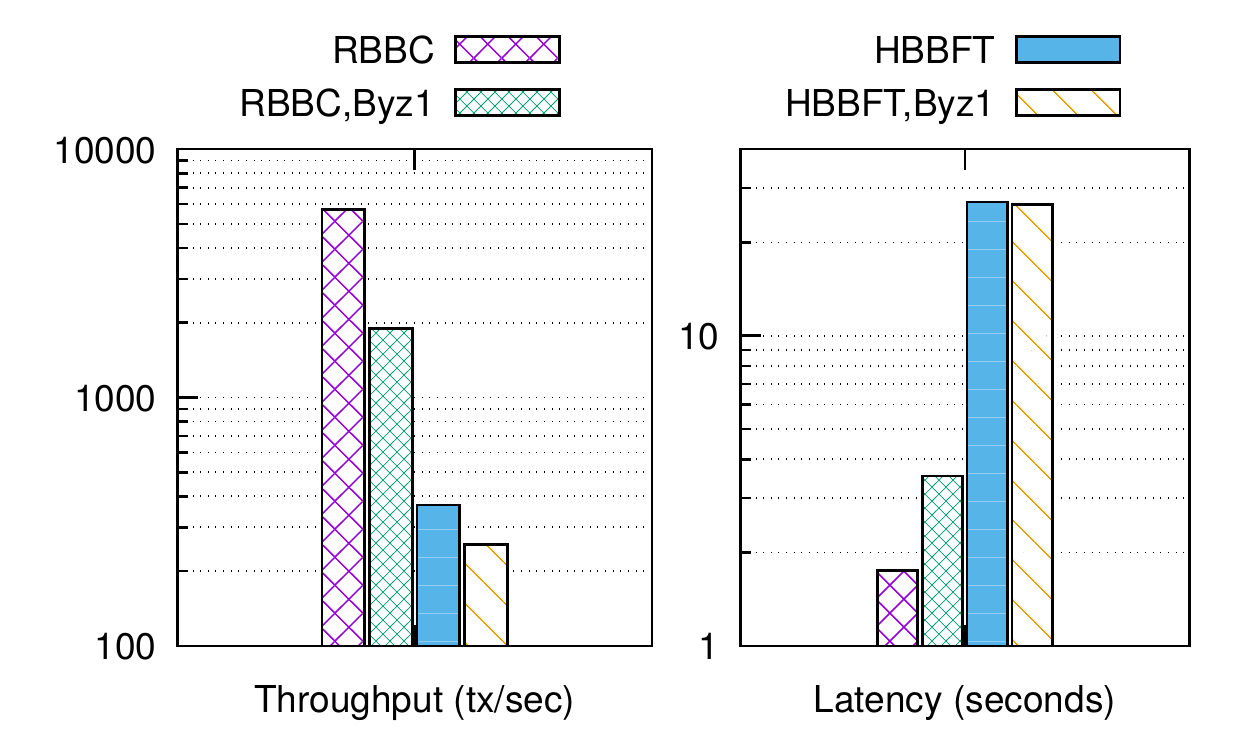}
         \caption{Comparing throughput and latency of RBBC and HBBFT, with normal and Byzantine behavior on 100 geodistributed  nodes; all $n$ nodes are making proposals of 100 transactions
           \label{fig:byz-tpt}}
       \end{center}
   \vspace{-2em}
\end{figure}

Figure~\ref{fig:byz-tpt}
shows the impact of Byz1 on performance with $n$ proposers and proposal sizes of $100$.
For RBBC, throughput 
drops from 5700\,tx/sec to 1900\,tx/sec, and latency increases from 920\,ms to 1750\,ms.
The drop in throughput is partially due to having $t$ less proposals being accepted (the proposals
sent by Byzantine nodes are invalid), and to the increase in latency.
The increase in latency is 
due to the extra rounds needed to be executed
by the binary consensus to terminate with $0$.
The throughput of HBBFT drops from 350 to 256\,tx/sec due to the decrease in proposals, but interestingly the latency also decreases.
This is due to the fact that since there are less proposals, less computation is needed for the erasure codes.

\begin{figure}[t]
       \begin{center}
         \includegraphics[scale=0.65]{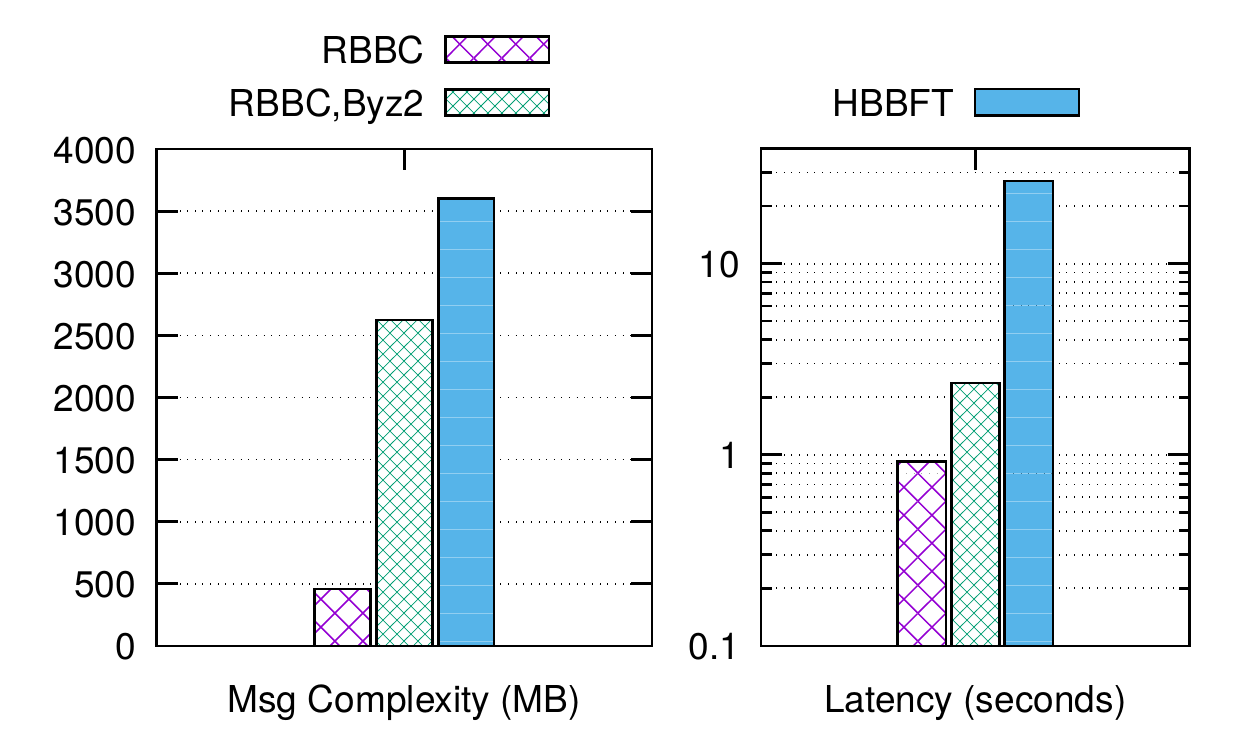}
         \caption{Comparing message complexity and latency of RBBFT and HBBFT with normal and Byzantine behaviors on 100
           geodistributed nodes
           \label{fig:byz-msg}}
       \end{center}
   \vspace{-1em}
\end{figure}

Byz2 is a behavior designed against the digest compression of the reliable broadcast,
with the goal of delaying the delivery of the message to $t$ of the correct nodes,
and increasing the bandwidth used.
HBBFT avoids this problem by using erasure codes, but has a higher bandwidth usage is the non-faulty case.
Figure~\ref{fig:byz-msg} shows the impact of this behavior on bandwidth usage and latency for RBBC and HBBFT with $n$ proposers and
proposal sizes of $100$.
The bandwidth usage of RBBC increases from 538\,MB per multivalued consensus instance to 2622\,MB per multivalued consensus instance
compared to HBBFT which uses 3600\,MB in all cases.
Furthermore, the latency of RBBC increases from 920\,ms to 2300\,ms.
Note that the bandwidth usage can further increase if additional delays are added to the network, in such cases
the use of Erasure codes becomes beneficial.

\begin{table*}[ht!]
  \begin{center}
   {\small 
   \arrayrulecolor{black}
\setlength{\tabcolsep}{10pt}
\begin{tabular}{ l  r r r r r r r }
& \#Requesters & Valid-tx/sec & Read/sec & R/W ratio & Latency(ms) &Valid-tx/block &  Invalid-tx/block \\
  \toprule 
  \multirow{4}{*}{RBBFT}
  & 1,000 & 5,359 & 2,143 & 0.4 & 870 & 4,648 & 0  \\
  & 10,000 & 13,870 & 33,288 & 2.4 & 2,475 & 34,132 & 877 \\
  & 20,000 & 12,664 & 31,660 & 2.5 & 5,022 & 63,607 & 3,033 \\
  & 50,000 & 14,450 & 47,685 & 3.3 & 4,303  & 62,193 & 5,455 \\
  \hline
  \multirow{4}{*}{CONS1}
  & 1,000 & 3,759 & 1,127  & 0.3 & 401 & 1,513 & 0 \\
  & 10,000 & 3,309 & 6,278 & 1.9 & 359 & 1,172 & 0 \\
  & 20,000 & 4,064 & 10,566 & 2.6 & 488 & 1,981 & 0 \\
  & 50,000 & 4,035 & 12,509 & 3.1 & 625 & 2,500 & 0 \\
  \bottomrule 
\end{tabular}
    }
    \caption{Performance of RBBFT and CONS1 with varying number of requesters.\label{fig:tab-cli-exp}}
  \end{center}
   \vspace{-2em}
\end{table*}

 \subsection{Single availability zone experiment}\label{sec:datacenter}
\begin{figure}[t]
         \begin{center}
         \includegraphics[scale=0.65]{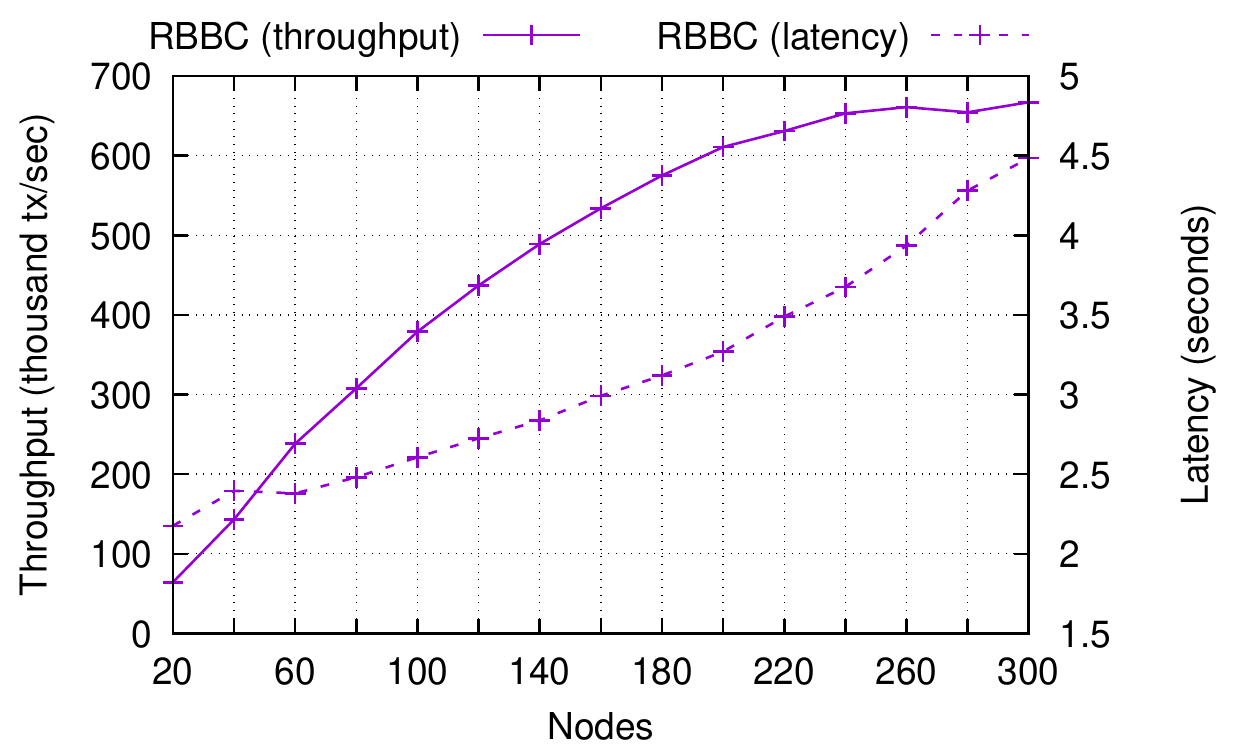}
         \caption{The performance (latency and throughput) of RBBC in a single datacenter\label{fig:scalability}}
         \end{center}
   \vspace{-2em}
 \end{figure}

To really stress test RBBC, we tested the performance on 300 
high-end VMs in the Oregon datacenter. 
We fixed $t$ to the largest fault tolerance parameter we can tolerate with $n=20$ nodes and increase the number of nodes from $20$ to $300$ permissioned nodes.
While the setting is not realistic, it helps identifying  potential performance bottlenecks. Note that Fig.~\ref{fig:consensus-fault-tolerance} depicts the impact of varying $t$ on performance.
 The results, shown in Figure~\ref{fig:scalability}, indicates that the throughput scales up to $n=260$ nodes to reach 660,000\,tx/sec while the latency remains lower than 4 seconds.
 At $n=280$ throughput drops slightly. Other experiments not shown here indicated about 
 $8$ verifications per transaction converging towards $7=t+1$ as $n$ increases.
 The performance is thus explained by the fact that the system is CPU-bound up to $n=260$, so that increasing $n$ adds CPU resources needed for the sharded verification and improves performance, after what the system becomes network-bound due to the consensus and performance flattens out. 

\begin{table*}[ht!]
   \begin{center}
    {\small 
     

\setlength{\tabcolsep}{8pt}
\begin{tabular}{ r r r r r r r  r }

  \toprule 
& \#Replicas & \#Requesters & Valid-tx/sec & Async write latency(ms) & Latency(ms) & Valid-tx/block &  Invalid-tx/block \\
\hline
& 1000 & 8400 & 30684 & 238 & 3103 & 95407 & 378   \\
\bottomrule
\end{tabular}
 }
     \caption{Performance of RBBFT with 1000 replicas spread in 14 data centers. \label{fig:tab-cli-exp2}}
   \end{center}
   \vspace{-2em}
\end{table*}

\subsection{Impact of remote requesters}\label{ssec:remote-req}
For the following experiments we run the blockchain with requesters
defined as follows.
At the start of the benchmark each requester is assigned a random private key
and a single UTXO contained within the genesis block with value 100,000 coins.
The requester then loops over the following two steps until the benchmark completes:
\emph{(i)}~For each UTXO currently assigned to the requester a new transaction is created
using that UTXO as input.
For the transaction's output a UTXO is created using a randomly chosen
account 
as the receiver with a value of 10 coins.
Any change is included in a second UTXO sent back to the requester.
Each transaction is then broadcast to the requester's assigned proposers.
\emph{(ii)}~The requester then repeatably performs the $\sf{request\_utxos}(\ms{account})$
operation until it receives at lest one new UTXO and then returns to step (i).
Each requester is run in its own thread and maintains connections to $2t+1$ of the blockchain nodes,
including the requester's $t+1$ proposers (all CONS1 requesters have the same
primary proposer).

For this experiment we ran RBBC and CONS1 using 100 c4.4xLarge server instances 25 c4.4xLarge requester instances. Both types of nodes are evenly distributed across
the 5 datacenters from US and Europe. 
The c4.4xLarge instances use Intel Xeon E5-2666 v3 processors with 16 vCPUs, and 30 GiB RAM.
The number of requesters vary from 1,000 to 50,000 and are evenly distributed across the requester nodes.
For the proposal size $\beta$, we choose $1000$ for RBBFT as it gave the best throughput.
For CONS1 we chose a proposal size of $2500$ as we found that larger sizes increased latency without increasing throughput
and smaller sizes decreased throughput while only having a minor impact on latency.
The experiments were run for 45 sec with a 15 sec warmup.

Table~\ref{fig:tab-cli-exp} shows the results.
\emph{Valid-tx/sec} is average number of valid transactions committed per second,
\emph{Read/sec} is the average number of $\sf{request\_utxos}(\ms{account})$ operations performed per second, \emph{R/W ratio}
is the ratio of the previous two values, \emph{Latency} is the average amount of time between committed blocks,
\emph{Valid-tx/block} is the average number of valid transactions per block, and \emph{Invalid-tx/block} is the average
number of invalid transactions per block.

Similar to the previous experiments we see that RBBFT has the highest maximum throughput of $14,450$ tx/sec compared to $4,064$ with CONS1.
RBBFT has the highest maximum latency between blocks of $5,022$ milliseconds compared to a maximum of $625$ milliseconds for CONS1.
The higher throughput and latency is explained by the higher utilization of resources by the sharded proposers and reduced computation
needed for sharded verification.
In RBBFT increasing the number of requesters past 10,000 has little impact on the throughput as the system resources are already saturated
by this point, as a result we see an increase in the R/W ratio as it takes longer for each individual node's transaction to complete.
A similar pattern is shown by CONS1, though this starts at 1,000 requesters as they are limited by the single primary proposer.
Furthermore in RBBFT, increasing the number of requesters also increases the number of duplicate transactions occurring in blocks.
This is due to the increased load in they system causing slower nodes to miss their proposals resulting in transactions being
committed by secondary proposers.

\section{Evaluation with 1000 machines}\label{ssec:thousand-vm}
To confirm that our blockchain scales to a large number of machines, we spawned 1000 VMs. 
To avoid wasting bandwidth, we segregated the roles: all 1000 VMs act as servers, keeping a local copy of the balances of all accounts. On these replicas, 10 requesters per 840 c4.large machines (60 VMs in each of 14 datacenters) send transactions and 160 c4.8xlarge machines (40 machines in each of the Ireland, London, Ohio and Oregon datacenters)
decide upon each block. 

Each of the 8400 requesters start with 100 UTXOs and each proposal contains up to 1000 transactions.
Performance are depicted in Table~\ref{fig:tab-cli-exp2}: throughput is only around 30,000 tx/sec due to the difficulty of generating the workload: the replicas are located in 14 different datacenters and have to wait for a UTXO to 
request a transaction that consumes it (cf. \cref{ssec:remote-req}). The asynchronous write latency measures the time a proposer acknowledges a transaction reception. The transaction commit time (latency) remains about 3 seconds despite the large traffic.

\section{Conclusion}\label{sec:conclusion}
In the most extensive experimentation of blockchain to date, we evaluated the Red Belly Blockchain, a deterministic blockchain system that does not need synchrony to be secure and performs well at large scale.
Its main novelty is its novel sharding that minimizes both computation and communication wastes that allows to achieve unprecedented throughput with a low latency when deployed world-wide.
The Red Belly Blockchain appears as a platform of choice for obtaining the security needed to move blockchain use-cases from innovation labs to production without sacrificing performance.

\bibliographystyle{abbrvnat}
\bibliography{reference} 

\end{document}